\documentclass[
  ,final            
  ]
  {aipproc}

\layoutstyle{6x9}

\begin{document}

\begin{flushright}
FERMILAB-CONF-11-381-E
\end{flushright}

\title{Studies of Single Electroweak Bosons at the Tevatron}


\author{Adam L. Lyon}{
  address={(For the CDF and D0 Collaborations)\\Fermi National Accelerator Laboratory, Batavia, IL 60510, USA}
}

\begin{abstract}
Tests of the Standard Model with Electroweak Physics have been
performed over mnay decades. In these proceedings, we present several
analyses from the Tevatron involving single $W$ or $Z$ bosons.
\end{abstract}

\maketitle


Electroweak (EW) physics has been crucial for discovering and
confirming many aspects of the Standard Model (SM). Furthermore,
through radiative corrections EW physics allows for indirect views of
heavy particles. Indeed the relationship between $W$ boson, top quark,
and Higgs boson masses is instrumental in predicting at what mass the
Higgs boson may finally be found. In these proceedings, we present several
recent analyses from the Tevatron involving single $W$ or $Z$ bosons
to test the SM.

The Tevatron is a $p\bar{p}$ collider at $\sqrt{s}  = 1.96$~TeV
located at the Fermi National Accelerator Laboratory.  CDF and D0 are
its two multi-purpose detectors concentrating on high $P_T$
physics. Both detectors are described in detail elsewhere\cite{cdf, d0}.

\subsection{$W$ boson mass and width}
Here, we merely summarize the Tevatron $W$ boson mass and width
results. D0 (CDF) has analyzed data corresponding to 1 (0.2) fb$^{-1}$
of luminosity. Figure 1 shows the combined Tevatron results in
comparison with other measurements for both the $W$ boson
mass\cite{wcombo} and width\cite{wwidth}. Both experiments are aiming
for future mass results with $\sim 25$~MeV$/c^2$ precision.

\begin{figure}[h]
\includegraphics[width=0.55\textwidth]{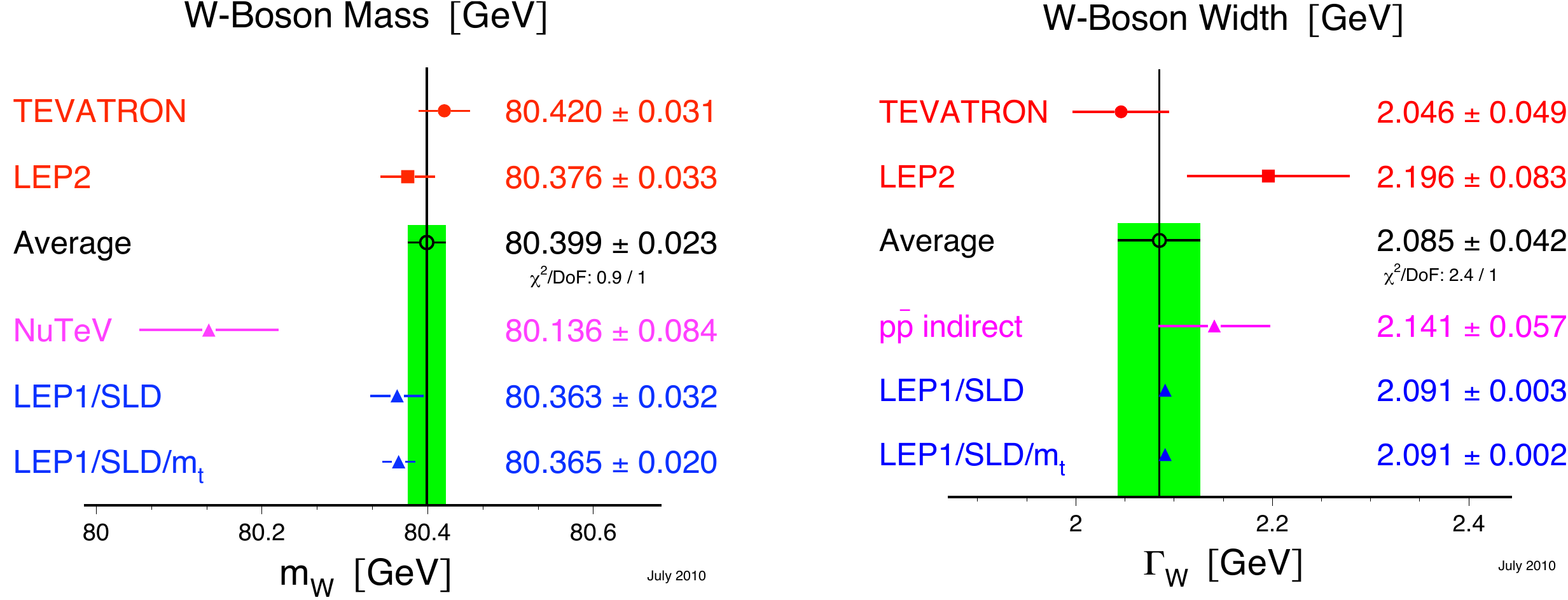}
\caption{Combined Tevatron results for the $W$ boson mass and width
  compared to other experiments.}
\end{figure}

\subsection{$Z$ boson rapidity}
The remaining sections of these proceedings involve properties of
$Z$ boson production. Note that $Z$ includes not only on shell $Z$
bosons, but also Drell-Yan events from a produced $\gamma^*$. The
first such analysis we consider is a measurement of the $Z$ boson
rapidity $d\sigma /dy$ from decays to $ee$ pairs (daughters of
$Z/\gamma^*$ are always oppositely charged pairs) by CDF\cite{dsigdy}.
Events at large rapidity are produced in collisions where the momentum
fraction $x$ of the quarks in either the proton or antiproton is
large. Therefore, such events provide tests of parton distribution
functions (PDFs) at large $x$. Data corresponding to 2.1~fb$^{-1}$
were utilized.

Figure 2 displays the measured unfolded distribution and then
comparisons to several PDFs.  One notices that the next-to-leading
order (NLO) CTEQ6 PDFs agree quite well with the data and that the
next-to-next-to-leading order (NNLO) MSTW2008 agrees better than the
NLO version.  Determining the inclusive cross section gives  $256.6 \pm
0.7_{\mathrm{stat}} \pm 2.0_{\mathrm{sys}} \pm 15.4_{\mathrm{lum}}$~pb,
the most precise measurement of this quantity to date.

\begin{figure}
\includegraphics[width=0.60\textwidth]{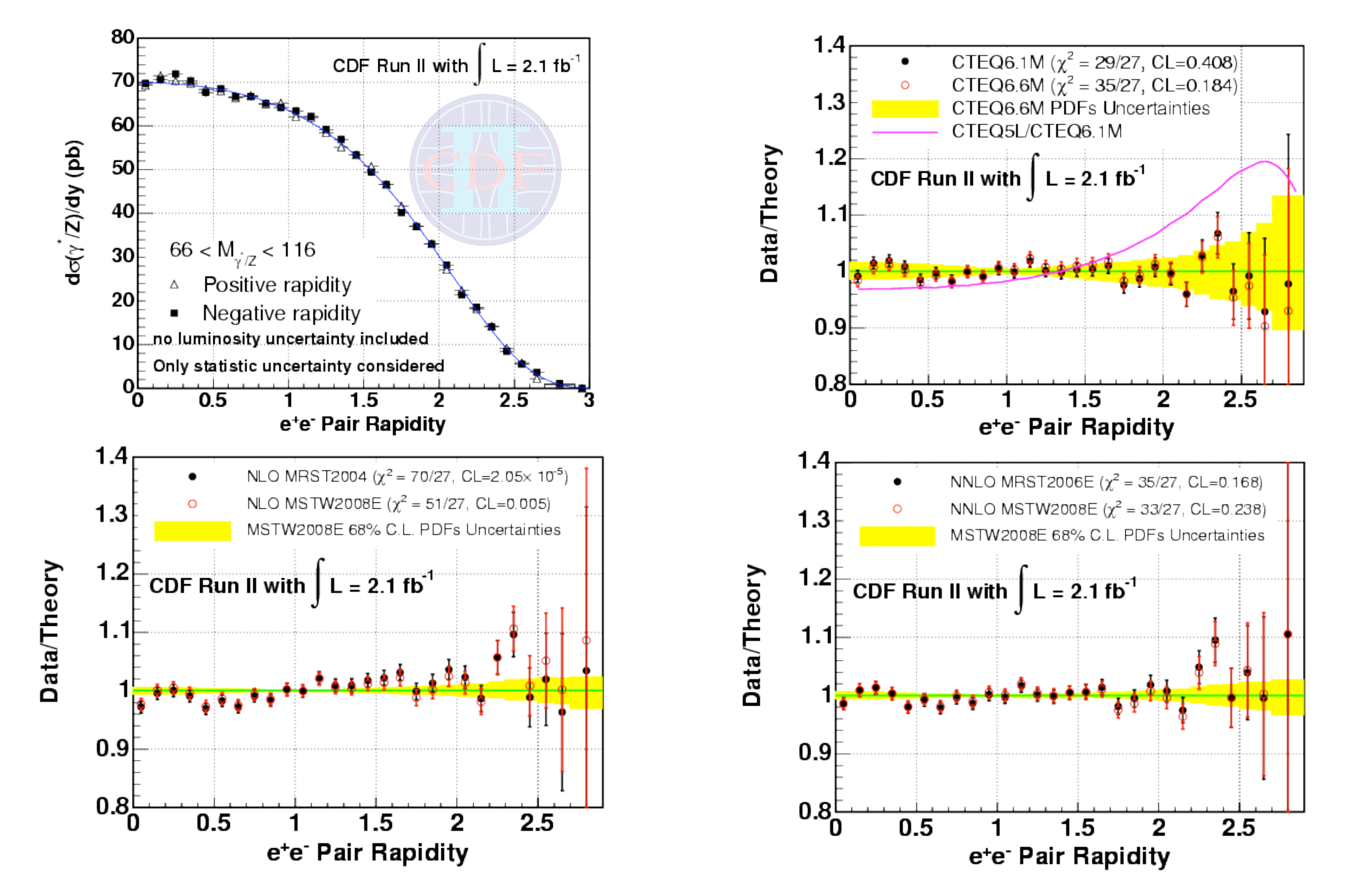}
\caption{Measurement of $Z/\gamma^*$  $d\sigma/dy$ from CDF and
  comparisons to PDFs.}
\end{figure}

\subsection{Tests of resummation and pQCD with Drell-Yan $P_T^{\mu\mu}$}

Leading order (LO) QCD corrections to the Drell-Yan production diagram
involve production of a $Z/\gamma^*$ with an additional gluon via
$q\bar{q}$ annihilation and production of $Z/\gamma^*$ with an
additional quark via $qg$ compton scattering. By measuring the $P_T$
of the $Z/\gamma^*$ one studies these corrections and tests
predictions from fixed order perturbative QCD (pQCD) at large $P_T$
and gluon resummation at low $P_T$. D0 analyzed data corresponding to
0.97~fb$^{-1}$ and measured the $Z/\gamma^*$ $P_T$ with $\mu\mu$
pairs\cite{d0zpt}. Figure 3 displays the measured spectrum corrected
for resolutions and efficiency only along with comparisons to various
event generators. One notes decent agreement except for generators
using tune D6. RESBOS includes gluon resummation and nicely matches
the low $P_T$ data. Nearly all of the generators have a normalization
problem when matching the data in the range 30 - 100~GeV/$c$.

\begin{figure}
\includegraphics[width=0.50\textwidth]{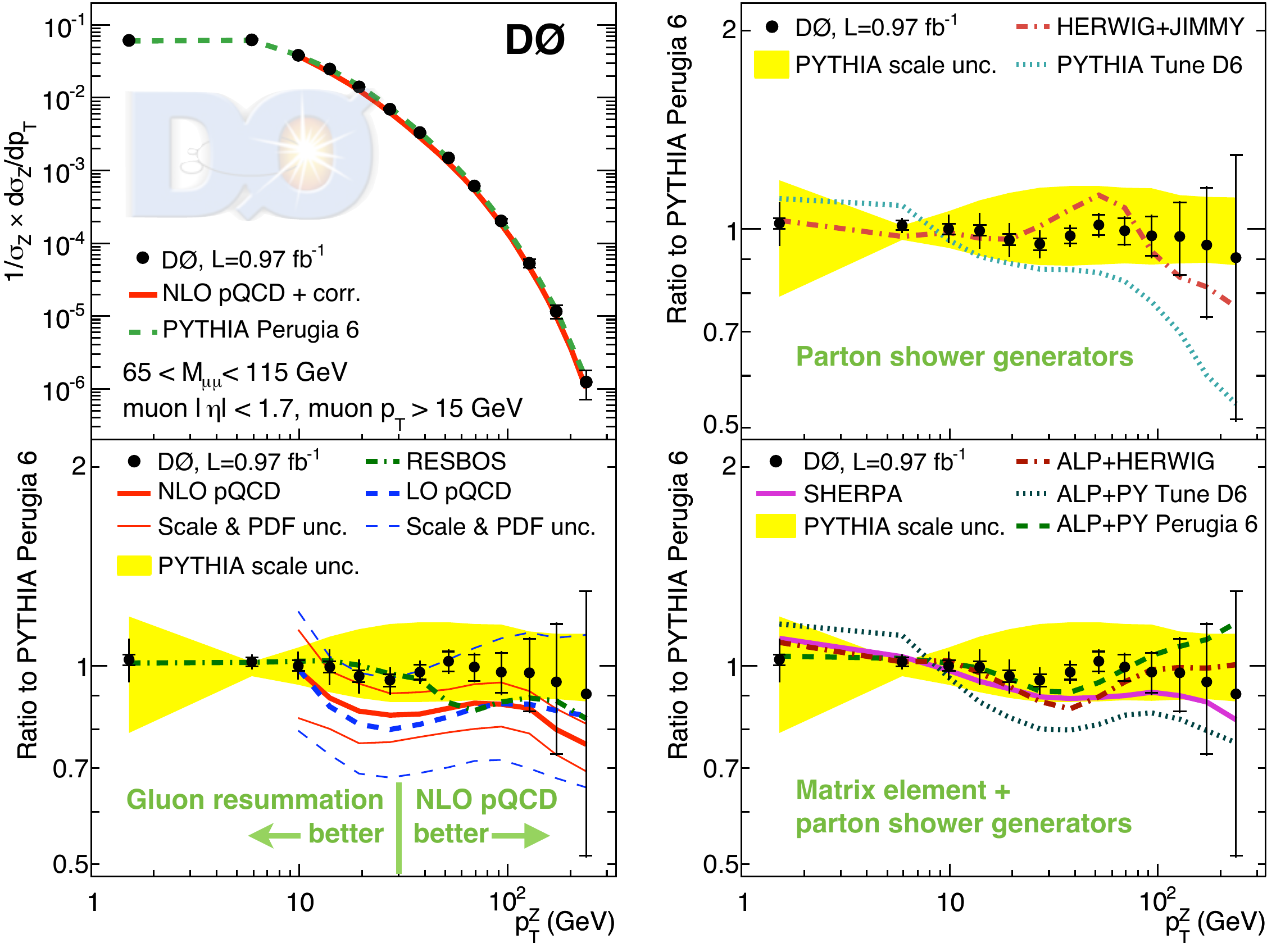}
\caption{Measurement of $Z/\gamma^*$  $P_T$ from D0 and comparisons to
  various generators.}
\end{figure}

\subsection{Improving $Z/\gamma^*$ $P_T^Z$ with angles}

Measuring the $Z/\gamma^*$ $P_T$ leads to several limitations. For low $P_T$, the
correction of the resolution and efficiencies dominates the
uncertainty. Furthermore, the bin widths of the distribution
are governed by resolution and not statistics.

Decomposing $P_T$ with respect to a thrust axis will lead to a
variable $a_T$ which is less susceptible to the problems mentioned
above\cite{at}, but further improvement may be gained by measuring
angles instead of momentum. An angle $\phi^*_{\eta}$ is constructed\cite{phistar}
with

\[ \phi^*_{\eta} = \tan (\phi_{\mathrm{acop}}/2) \sin(\theta^*_{\eta}) \]

\noindent where $\phi_{\mathrm{acop}}$ is an acoplanarity angle and
$\cos{\theta^*_{\eta}} = \tanh ((\eta^- - \eta^+)/2)$. $\phi^*_\eta$
is highly correlated with $a_T/m_{\ell\ell}$ (where the denominator is
the dilepton invariant mass) and has significantly improved resolution
due to the excellent angular resolution of the D0 tracking detectors.
In nearly all bins of angle, the total systematic uncertainty is
substantially smaller than the statistical uncertainty. Figure 4 shows
a comparison of the $\phi^*_\eta$ distribution in both the electron
and muon channels with RESBOS. Data corresponding to 7.3~fb$^{-1}$
were utilized. One notes the deficit at large angle, corresponding to
the similar deficit in the $P_T$ distributions, is evident.
Furthermore, RESBOS with small-$x$ broadening is disfavored.

\begin{figure}
\includegraphics[width=0.55\textwidth]{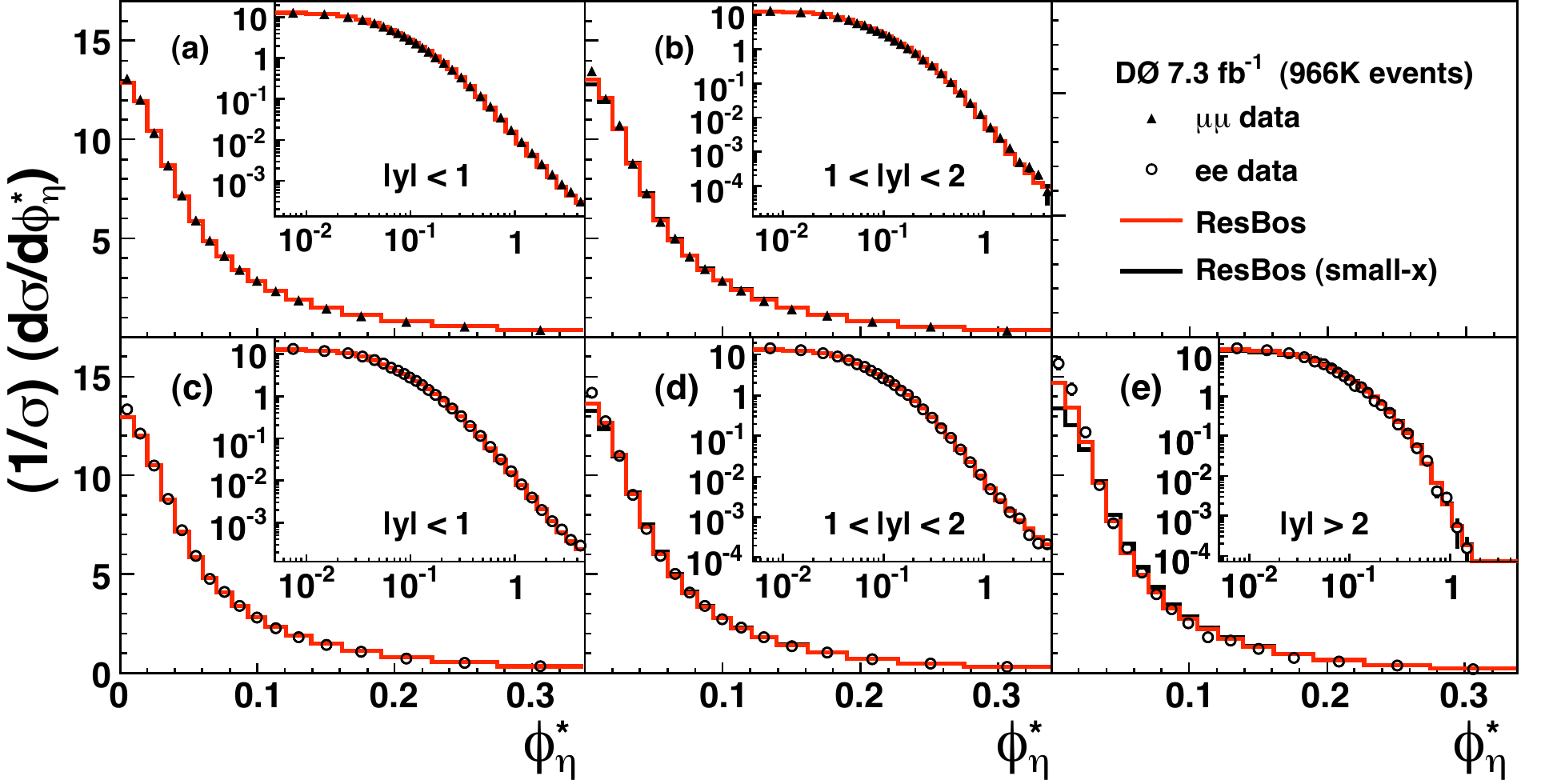}
\caption{Measurement of $Z/\gamma^*$  $\phi^*_{\eta}$  from D0 and comparisons to
  RESBOS.}
\end{figure}

\subsection{Tests of pQCD with Angular Coefficients of Drell-Yan $ee$
  pairs and $P_T^{ee}$}
This topic is covered in detail elsewhere in these
proceedings\cite{bodek}. The differential cross section $d\sigma / \ d
\cos\theta d\phi$ is expanded into terms involving angles in the
Collins-Soper center of mass frame. The coefficients are measured by
CDF\cite{cdfang} with data corresponding to 2.1~fb$^{-1}$ and compared
to pQCD predictions from various generators. Two interesting
by-products of this procedure are an indirect measurement of $\sin^2
\theta_W$ and the ``Lam-Tung relation'' which is sensitive to the spin
of the gluon.  $\sin^2 \theta_W$ is related to the $A_4$
coefficient. CDF thus measures $\sin^2 \theta_W \, = \, 0.2329 \pm
0.0008 ^{+0.0010}_{-0.0009}$ where the first uncertainty is from the
measurement of $A_4$ and the second is a systematic from QCD theory
and PDFs.  The ``Lam-Tung relation'' stipulates that coefficients $A_0
= A_2$ when the gluon has spin 1. CDF measures this difference in
coefficients across a range of $P_T^Z$ bins. The average difference is
$0.02 \pm 0.02$, consistent with a spin 1 gluon.

\subsection{Conclusion}
Several analyses involving single electroweak bosons have been
presented here, and all are important tests of the Standard Model. All
of these results are generally in good agreement with SM predictions,
and many of the comparisons will be used to fine-tune PDFs and
generators.



\bibliographystyle{aipproc}   


\end{document}